\documentclass{article}
\usepackage{arxiv}
\usepackage{indentfirst}
\usepackage[utf8]{inputenc} 
\usepackage[T1]{fontenc}    
\usepackage{hyperref}       
\hypersetup{
    colorlinks=true,
    linkcolor=blue,
    filecolor=magenta,      
    urlcolor=cyan,
}
\usepackage{url}            
\usepackage{booktabs}       
\usepackage{amsfonts}       
\usepackage{nicefrac}       
\usepackage{microtype}      
\usepackage{lipsum}
\usepackage[square,numbers]{natbib}
\usepackage{graphicx}
\usepackage[ruled,vlined]{algorithm2e}
\usepackage{amsmath}
\usepackage{float} 
\usepackage{ragged2e}
\usepackage{adjustbox}
\usepackage{caption}
\usepackage{xcolor,colortbl}
\usepackage{changepage}  
\usepackage[super]{nth}

\begin{document}

\renewcommand*{\thefootnote}{\arabic{footnote}}

\title{A Re-Examination of the Evidence used by Hooge et al (2018) \textit{``Is human classification by experienced untrained observers a gold standard in fixation detection?''}}

\author{
  Lee Friedman\\
  Department of Computer Science\\
  Texas State University\\
  San Marcos, Texas, USA, 78666\\
  \texttt{lfriedman10@gmail.com} \\
}

\date{Received: date / Accepted: date}

\maketitle

\begin{abstract}
Hooge et al. \cite{HoogeF1} asked the question: ``Is human classification by experienced untrained observers a gold standard in fixation detection?'' They conclude the answer is no.  If they had entitled their paper: ``Is human classification by experienced untrained observers a gold standard in fixation detection when data quality is very poor, data are error-filled, data presentation was not optimal, and the analysis was seriously flawed?'', I would have no case to make. In the present report, I will present evidence to support my view that this latter title is justified.  The low quality data assessment is based on using a relatively imprecise eye-tracker, the absence of  head restraint for any subjects, and the use of infants as the majority of subjects (60 of 70 subjects).  Allowing subjects with more than 50\% missing data (as much as 95\%) is also evidence of low quality data.  The error-filled assessment is based on evidence that a number of the ``fixations'' classified by ``experts'' have obvious saccades within them, and that, apparently, a number of fixations were classified on the basis of no signal at all.  The evidence for non-optimal data presentation stems from the fact that, in a number of cases, perfectly good data was not presented to the coders.  The flaws in the analysis are evidenced by the fact that entire stretches of missing data were considered classified, and that the measurement of saccade amplitude was based on many cases in which there was no saccade at all.  Without general evidence to the contrary, it is correct to assume that some human classifiers under some conditions may meet the criteria for a gold standard, and classifiers under other conditions may not.  This conditionality is not recognized by Hooge et al. \cite{HoogeF1}.  A fair assessment would conclude that whether or not humans can be considered a gold standard is still very much an open question.
\end{abstract}

\section{Introduction}
Hooge et al. \cite{HoogeF1} studied human classification of eye-tracking data.  They were specifically interested in the classification of fixations.  They concluded that human fixation classification cannot be considered a gold standard.  I recently had a close look at all of their evidence \footnote{The eyetracking data for \cite{HoogeF1} are available here: \color{blue} \url{https://doi.org/10.5281/zenodo.838313}\color{black}} for an unrelated reason.  What I saw surprised me.  In my view, some of the data and analysis was unsuitable to support such an important conclusion. I felt it was important that other researchers be made aware of this. 

In that paper, they provide minimal illustrations of their signals.  Specifically, they illustrate one second of data, from one subject, from a total of 352.2 seconds from 70 subjects (0.28\% of the recordings).  In the present report, I illustrate more than an order of magnitude more data than the original report.\footnote{To see all of the signals from the Hooge et al. (2018) study, as classified by rater MN, see https://digital.library.txstate.edu/handle/10877/9176. OriginalHoogeStyleFiguresAsPNG.zip are in .png form and OriginalHoogeStyleFiguresAsFIG.zip are in the MATLAB native .fig file form. }

\vspace{-15mm}\section{Method Details for Hooge et al. (2018)}

In describing their data, the authors state:

\begin{adjustwidth}{1cm}{}
``Twelve experienced but untrained human coders classified fixations in 6 min of adult and infant eye-tracking data.'' (page 1864)
\end{adjustwidth}

\noindent
\\
Within this manuscript, we will be exclusively employing the fixation classifications of a single coder, Marcus Nyström, (``MN'').  Dr. Nyström is an internationally recognized expert in eye movement classification.\\

The authors also state:

\begin{adjustwidth}{1cm}{}
``The eye-tracking stimulus set consists of 70 trials of eye-tracking data measured with a Tobii TX300 at 300 Hz. We used eye-tracking data measured from the left eye. Ten of the 70 trials contained 150.1 s of eye-tracking data of two adults looking at Roy Hessels’s holiday pictures taken in the arctic area around Tromsø, Norway. The other 60 trials contained 202.1 s of eye-tracking data of infants performing a search task \cite{HesselsInDepth}.'' (page 1867)
\end{adjustwidth}

Details of the scoring are also described:

\begin{adjustwidth}{1cm}{}
``Trials of both the adult and the infant eye-tracking datasets were presented in random order on a 24-in. TFT screen (1,920 $X$ 1,200 pixels). 
The vertical axis of the position signals was fixed (respectively, 0–1,920 and 0–1,080 pixels, since measurements were done on the HD screen of the TX300). 
...Each screen showed 1 s of data and contained the last 250 ms of the previous display (to provide context) and 750 ms new data at a time.'' (page 1867)
\end{adjustwidth}

In a personal communication with Dr. Hooge, I was informed that:``We did not use a forehead and chinrest.''

Also, I made the following observations of the data: Adult data consisted of approximately 4500 samples (range: 4499 - 4504, approximately 13.5 seconds). Most infant recordings (46/60) consisted of approximately 1200 samples (range: 1203  - 1210, approximately 3.6 seconds), but 14/60 recordings had very short recordings (median of 312.5 samples, approximately, 0.94 seconds) with a range of 157 - 769 samples.  None of these 14 short recordings are plotted in the present report.

\vspace{-15mm}\section{Results}
\vspace{-7mm}\subsection{Examples of Good Recordings}

Some of the adult and infant recordings looked very reasonable (Figures 1 and 2).  Generally, the data from adults looked reasonable.  As we will see, the infant data are  more problematic.

\quad\quad\quad
\frame{\includegraphics[width=15cm]{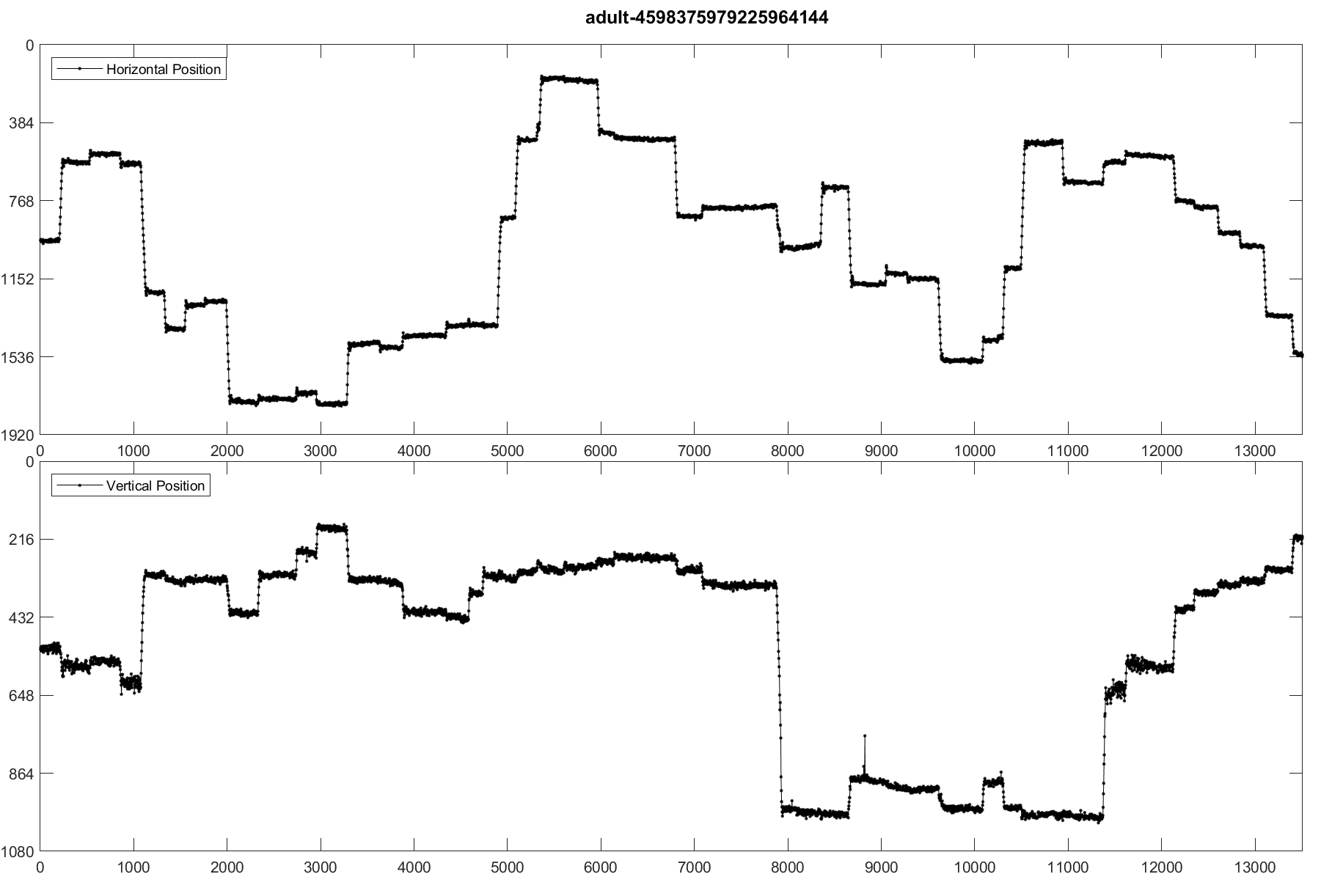}}
\begin{picture}(500,65)(0,0) \put(0,5){{\parbox{500\unitlength}{
{\textbf{Fig. 1. This is a good recording from an adult subject (adult-4598375979225964144).} The top panel represents horizontal position, in pixels. It is scaled from 0 to 1,920, which matches the scaling \cite{HoogeF1} used for the hand classification task.  The x-axis is scaled in milliseconds. The bottom panel represents vertical position, in pixels. It is scaled from 0 to 1,080, which also matches the scaling used for the hand classification task.  Both plots represent approximately 13.5 seconds of data.}
}}}
\end{picture}

\quad\quad\quad\quad
\frame{\includegraphics[width=0.8\textwidth]{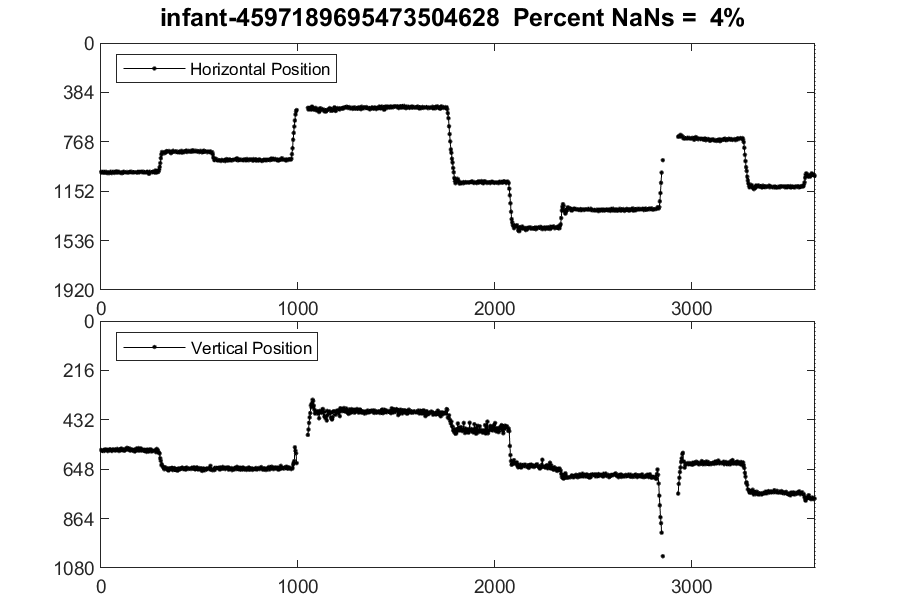}}
\begin{picture}(500,55)(0,0) \put(0,5){{\parbox{500\unitlength}{
\textbf{Fig. 2. This is a good recording from an infant subject.} It plots approximately 3.6 seconds of data. The infant recordings all indicate the percent of the recording that is missing (NaN), for reasons that will be clear below. See caption for Figure 1.}
}}
\end{picture}

\clearpage
\subsection{Out-of-Range Values}
Human classification of fixations is likely to be better when all relevant information is provided to the rater.  
As noted above, for the human classification of these recordings, the eye position was represented in pixels.  The horizontal position signal was scaled to go from 0-1920 pixels and the vertical position signals was scaled from 0-1080. I have found that, in some cases, a substantial amount of the signal, and in particular, the vertical position signal, is out of the presented range. Table 2 lists all of the studies where more than 100 samples were out-of-range. Figure 3 (below) illustrates 4 such cases.  By clipping the data ranges, Hooge et al. have not optimally presented their data to raters.

\begin{table*}[b]
\center
\caption{List of Studies with More than 100 Out-of-Range Samples}
\begin{tabular}{|c|c|}
\hline
Data Set & Number of \\ & Out-of-Range \\
& Samples \\ \hline
infant-4601345468058933046 & 639 \\ \hline
infant-4606386303357509649 & 539 \\ \hline
infant-4597253268938977956 & 528 \\ \hline
infant-4603602825832748254 & 411 \\ \hline
infant-4595441564971508404 & 341 \\ \hline
infant-4606604723864184111 & 313 \\ \hline
infant-4604530757509485450 & 307 \\ \hline
infant-4595343491140989604 & 275 \\ \hline
infant-4597178908827295394 & 235 \\ \hline
infant-4584726578059882688 & 218 \\ \hline
infant-4605631479618949368 & 212 \\ \hline
infant-4604305481598970292 & 157 \\ \hline
infant-4595653501593606004 & 103 \\ \hline
\end{tabular}
\end{table*}
\clearpage

\quad\quad
\frame{\includegraphics[width=0.9\textwidth]{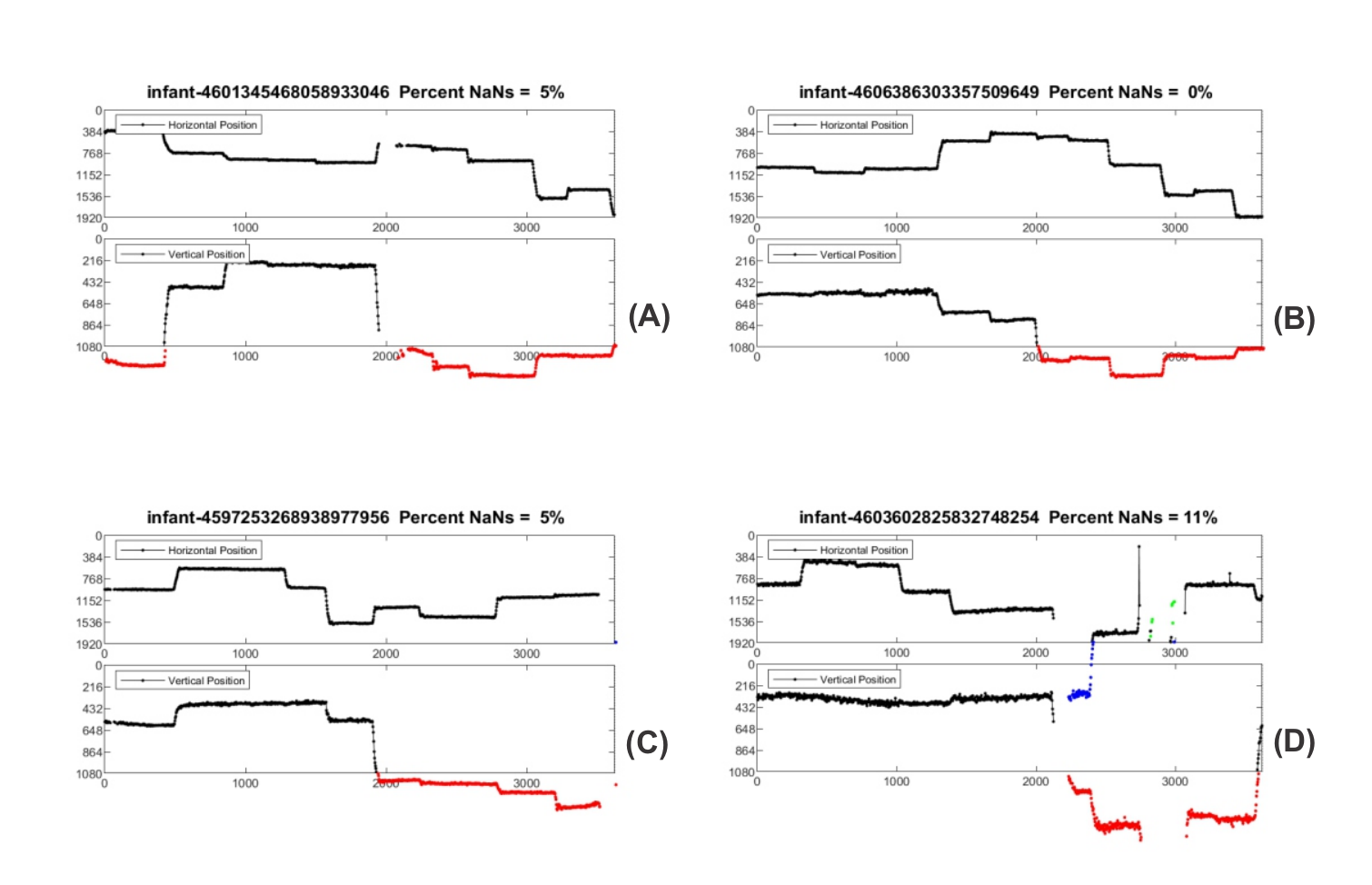}}
\begin{picture}(500,35)(0,0) \put(0,5){{\parbox{500\unitlength}{
\textbf{Fig. 3: Four examples of infant recordings with out-of-range values.}  The red dots indicate where vertical position data is larger than 1080. In D, the blue dots represent horizontal positions larger than 1,920. The green dots represent data from the vertical channel that is less than 0.0, and thus out-of-range.  The red signal and the blue signal appear to be actual eye movements and not artifact.  Visualization of this data would provide a human rater with some useful additional information. }
}}
\end{picture}

\subsection{Infant Recordings with Missing Data}

A large number of these recordings have a great deal of missing data.  Table 2 lists the 19 of 60 infant recordings with more than 25\% missing data.
Figure 4 illustrates the 4 infant recordings with more than approximately 83\% missing data.

\begin{table*}[htbp]
\center
\caption{List of Subjects More than 25\% NaN Values}
\begin{tabular}{|c|c|c|c|}
\hline
FileName & Number of NaNs & Percent of NaNs \\ \hline
infant-4605026287769552265 & 1135 & 94.347 \\ \hline
infant-4601213487207749474 & 1075 & 89.212 \\ \hline
infant-4607024019025466472 & 1009 & 83.874 \\ \hline
infant-4606050638313716974 & 1002 & 82.947 \\ \hline
infant-4603997294600609666 & 955 & 79.253 \\ \hline
infant-4607158533597356135 & 933 & 77.427 \\ \hline
infant-4597078871350441716 & 842 & 69.934 \\ \hline
infant-4585006770229471040 & 840 & 69.825 \\ \hline
infant-4590914421796899792 & 746 & 61.96 \\ \hline
infant-4604414452347604331 & 699 & 58.008 \\ \hline
infant-4599348903986769270 & 639 & 53.073 \\ \hline
infant-4603403341432416855 & 514 & 42.62 \\ \hline
infant-4606305162693909260 & 511 & 42.442 \\ \hline
infant-4592207091392917904 & 491 & 40.612 \\ \hline
infant-4591350692076781768 & 455 & 37.759 \\ \hline
infant-4605410513238776168 & 372 & 30.846 \\ \hline
infant-4606003927763488057 & 364 & 30.258 \\ \hline
infant-4604305481598970292 & 342 & 28.288 \\ \hline
infant-4605849449150400922 & 315 & 26.033 \\ \hline
\end{tabular}
\end{table*}

\quad\quad\quad\quad
\frame{\includegraphics[width=0.8\textwidth]{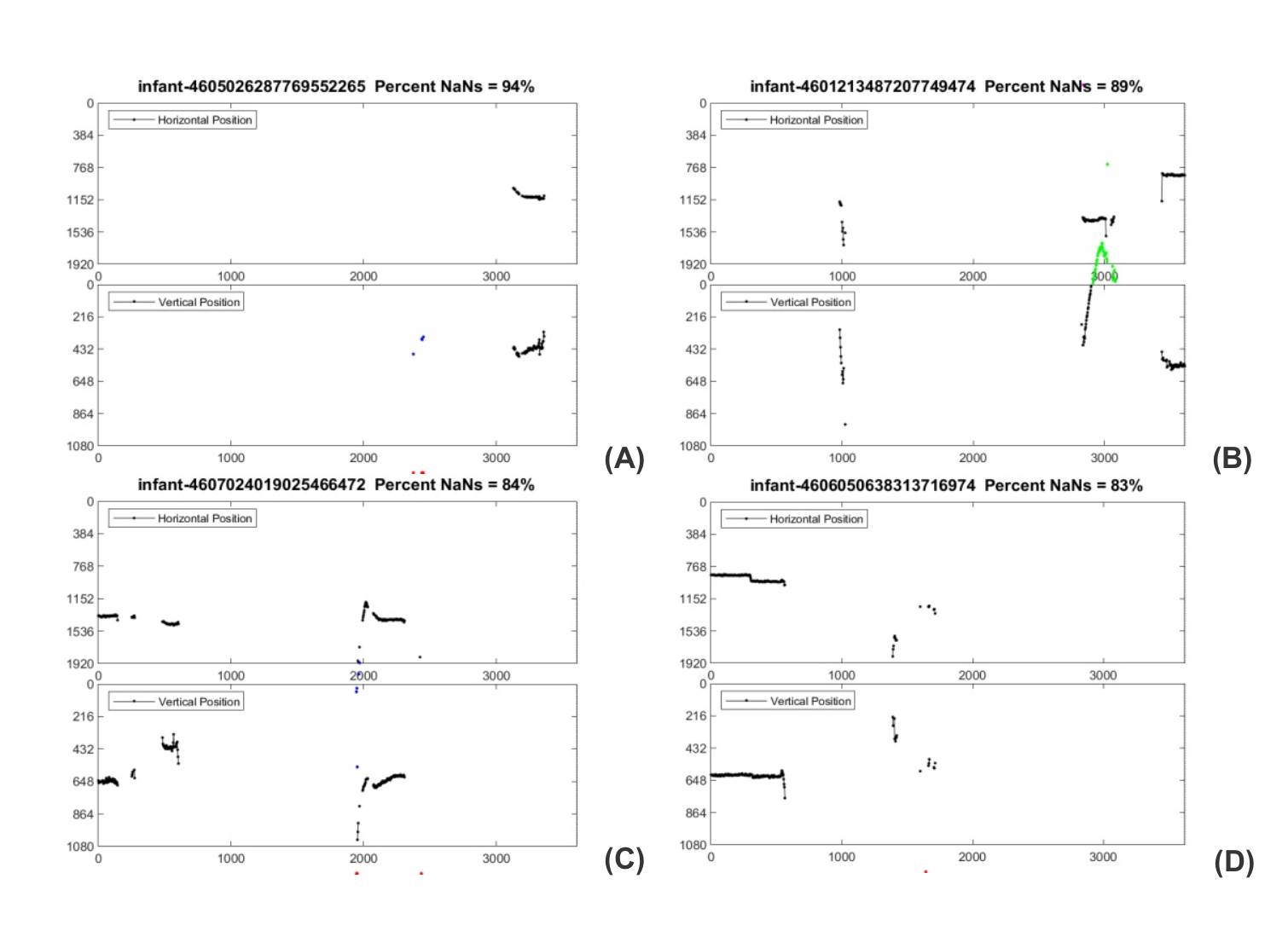}}
\begin{picture}(500,20)(0,0) \put(0,5){{\parbox{500\unitlength}{
\textbf{Fig. 4: Four infant recordings with between 83\% to 94\% missing data.}}
}}
\end{picture}

\subsection{Missing Data Classified as Fixation}
There are numerous examples where missing data were classified as fixation (see Figs. 6-9)

\vspace{20pt}
\quad\quad\quad\quad
\frame{\includegraphics[width=0.8\textwidth]{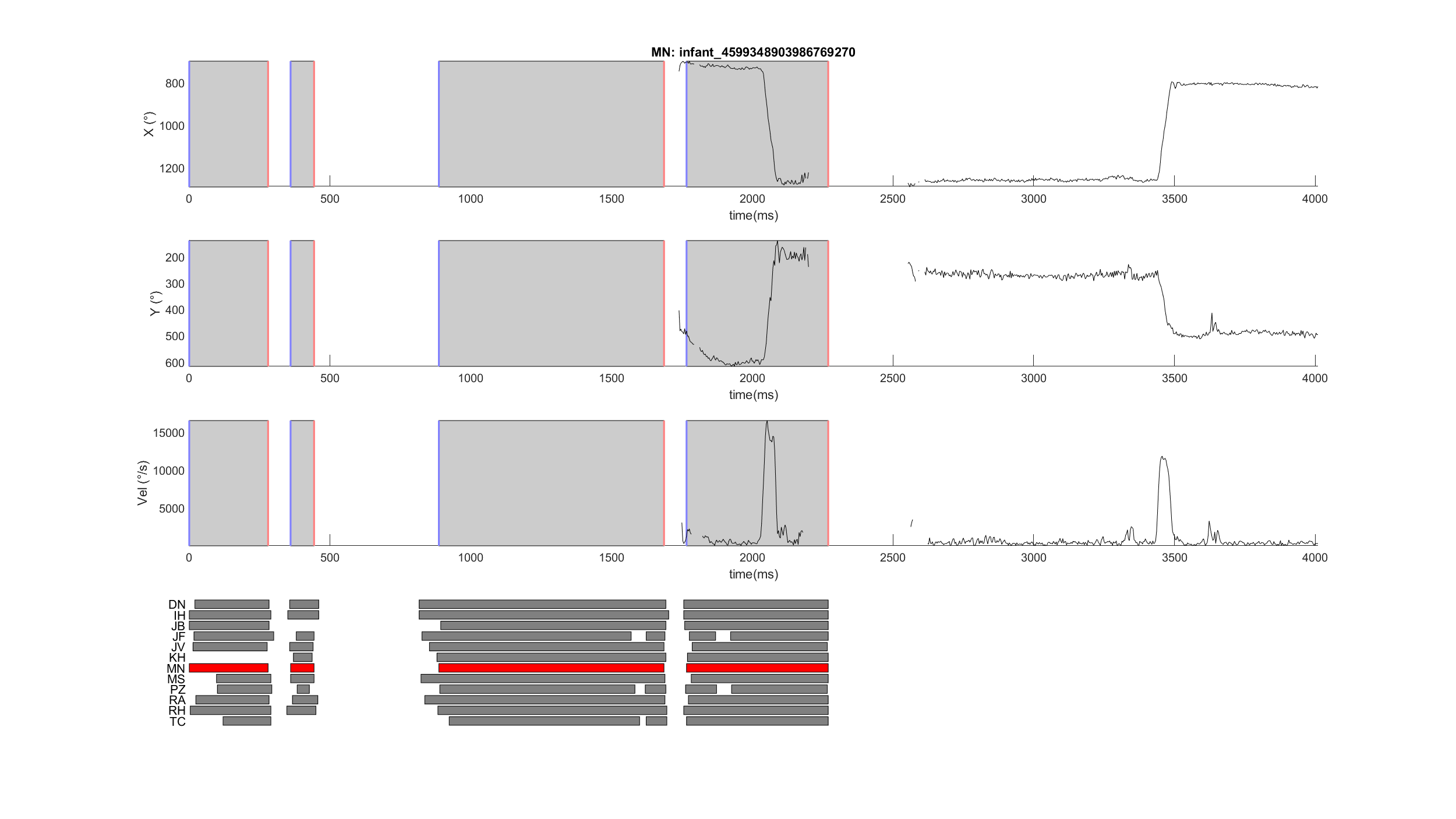}}
\begin{picture}(500,20)(0,0) \put(0,5){{\parbox{500\unitlength}{
\textbf{Fig. 6: Recording where missing data is classified as fixation.  Gray boxes in top 3 subplots indicate where rater MN classified fixation. Note the absence of a signal in these sections.}}
}}
\end{picture}


\clearpage
\quad\quad\quad\quad
\frame{\includegraphics[width=0.8\textwidth]{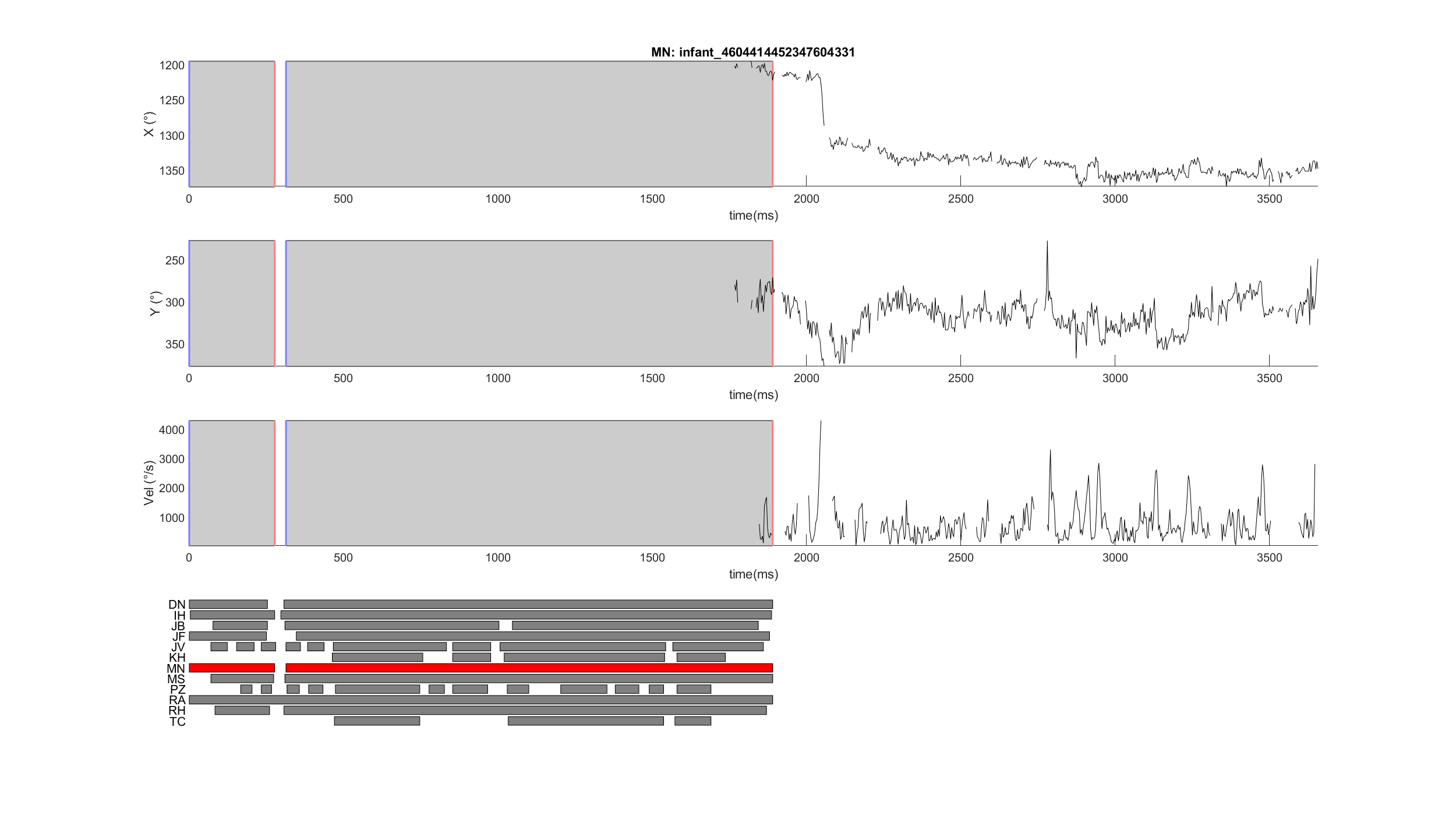}}
\begin{picture}(500,20)(0,0) \put(0,5){{\parbox{500\unitlength}{
\textbf{Fig. 8: Recording where missing data is classified as fixation. Gray boxes in top 3 subplots indicate where rater MN classified fixation. Note the absence of a signal in these sections.}}
}}
\end{picture}


\clearpage
\subsection{Inclusion of missing data in evaluation of agreement statistics}
We consider it axiomatic that humans cannot classify signals they cannot see.  However, Hooge et al. \cite{HoogeF1} include sections of missing data in their analysis of agreement between raters.  In other words, the considered that sections of recordings with missing data were still classifiable by human raters.   There is some sense to this, in that it allows raters to classify long fixations as fixations, even if broken up by a small amount of missing data.  However, it does not seem reasonable to me that an event can be considered classified if a majority of the event, or even the entire event, consists of missing data. In this section, we provide examples where an entire non-fixation event consisted of missing data.

\quad\quad\quad\quad
\frame{\includegraphics[width=0.8\textwidth]{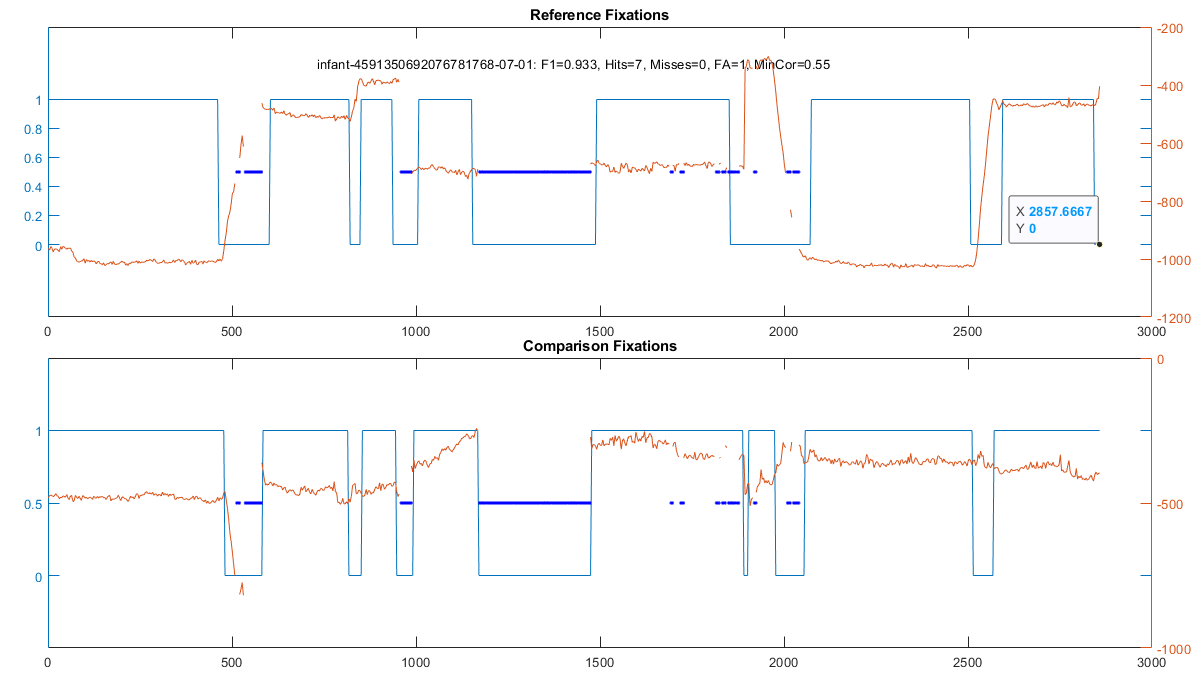}}
\begin{picture}(500,65)(0,0) \put(0,5){{\parbox{500\unitlength}{
\textbf{Fig. 10: These are plots of the data submitted for assessment of agreement statistics. The actual signals are plotted in red, with the y-axis in pixels to the right of each subplot.  The binary classification of fixation and non-fixation is shown in light blue lines.  In the upper subplot, the reference fixations are illustrated.  A value of 1 means that the reference rater classified fixation and a value of 0 means that the reference rater did not classify fixation.  The bottom plot illustrates the classification of the comparison rater.  The darker blue points indicate missing data.  Note the inclusion of stretches of missing data in the data considered classified.  For example, look at the middle of the lower subplot, where a long blue horizontal line appears to end near the 1500 tick mark.  This whole section is considered to be marked as non-fixation by the comparison rater.  However, the comparison rater could not have marked this as fixation or non-fixation since all the data are missing.  In this plot, there are two other non-fixation periods that appear to consist of a majority of missing data.}}  
}}
\end{picture}

\clearpage
\quad\quad\quad\quad
\frame{\includegraphics[width=0.8\textwidth]{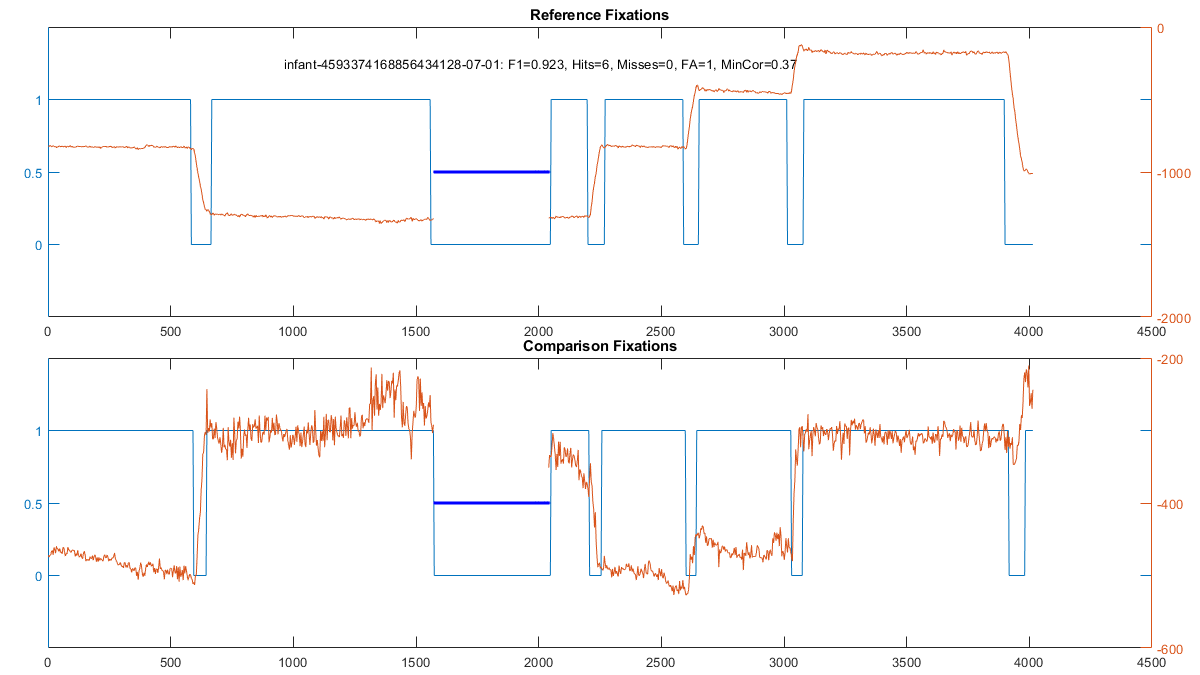}}
\begin{picture}(500,33)(0,0) \put(0,2){{\parbox{500\unitlength}{
\textbf{Fig. 11: These are plots of the data submitted for assessment of agreement statistics.  See caption for Fig. 10.  Note the inclusion of a stretch of missing data in the data considered classified.  See the lower subplot, where a long dark blue horizontal line appears to end near the 2000 tick mark.  This whole section was  considered to be marked as non-fixation by the comparison rater.  However, the comparison rater could not have marked this as fixation or non-fixation since all the data are missing.}}
}}
\end{picture}
\vspace{5mm}\subsection{Saccade-Based Evidence}

In \cite{HoogeF1}, the authors state: 

\begin{adjustwidth}{1cm}{}
''We asked the coders to mark fixations and not saccades, however between the majority of the fixations, saccade candidates are located. To find these saccade candidates we took periods of data between fixations with durations shorter than 100 ms and no data loss. This duration criterion is a liberal one; large 30 degree saccades last about 100 ms \cite{Collewijn}. From here on, we will refer to these intervals as saccade instead of saccade candidate.'' (page 1881)
\end{adjustwidth}

Within this manuscript, we will refer to these inter-fixation intervals as potential saccades.  The authors measure the amplitudes of these potential saccades, although the details of this measurement were not provided.  For example, each potential saccade had a horizontal and a vertical component.  How were these combined into a single estimate of saccade amplitude?  How was the amplitude actually determined: did they take the difference between the maximum and minimum amplitude within each inter-fixation interval as saccade amplitude? (We assume that this is what they did.) They argue that disagreements between human coders in terms of number of saccades and saccade amplitude  is further evidence that human classification is not a gold standard. I had a close look at these potential saccades, i.e., inter-fixation intervals as coded by rater MN. I used the same criteria as above (durations shorter than 100 ms and no data loss).  (I included such events that ended the recording.)

The first thing to note is that the overwhelming number of these saccade candidates did not resemble saccades whose onset and offset were marked by human experts.  In my opinion, there were only 14 of these 539 events whose starts and ends would be would be marked by an eye-movement expert as such.  Obviously, different raters will have different views. The readers are  encouraged to try this themselves. We present plots of all 539 saccades\footnote{See ``BadSaccades.pdf'', ``SaccadesAmplitudeOK.pdf'' and ``SaccadesAmplitudeNotOK.pdf at https://digital.library.txstate.edu/handle/10877/9176}.  But there is no doubt that the vast majority of the onsets and offsets of these events would not be marked by experts as the start and end of a saccade event.

However, Hooge et al \cite{HoogeF1} only use the amplitude of these events to provide evidence that human raters are not a gold standard.  Therefore, we classified each saccade into one of three groups: (1) bad saccades, i.e., no real evidence of a saccade in the interval, (2) saccade candidates whose amplitude would be reasonably determined based on the minimum and maximum amplitude change in the interval, and (3) saccades whose amplitude would not be reasonably determined based on the minimum and maximum amplitude change in the interval.  For this we measured the amplitude of each saccade and computed a ratio of that amplitude to the amplitude of the entire interval. If the ratio was >= 0.8, we considered this to be a reasonable amplitude measure.  If anything, we think this is over-generous.

There were 47 bad saccades.  Eight examples are presented in Figure 12.  So 8.7\% of ``saccades'' used by Hooge et al. \cite{HoogeF1} for rater MN (See their Fig. 5A) were not saccades. There were 11 whose amplitude would not be accurately measured if one used the maximum and minimum amplitude difference in these inter-fixation intervals.  Six examples are shown in Figure 13.  Thus, at the least, approximately 11\% of all of the saccade (55/539) amplitudes for rater MN (See their Fig. 5B) were not accurate.

\frame{\includegraphics[width=1.0\textwidth]{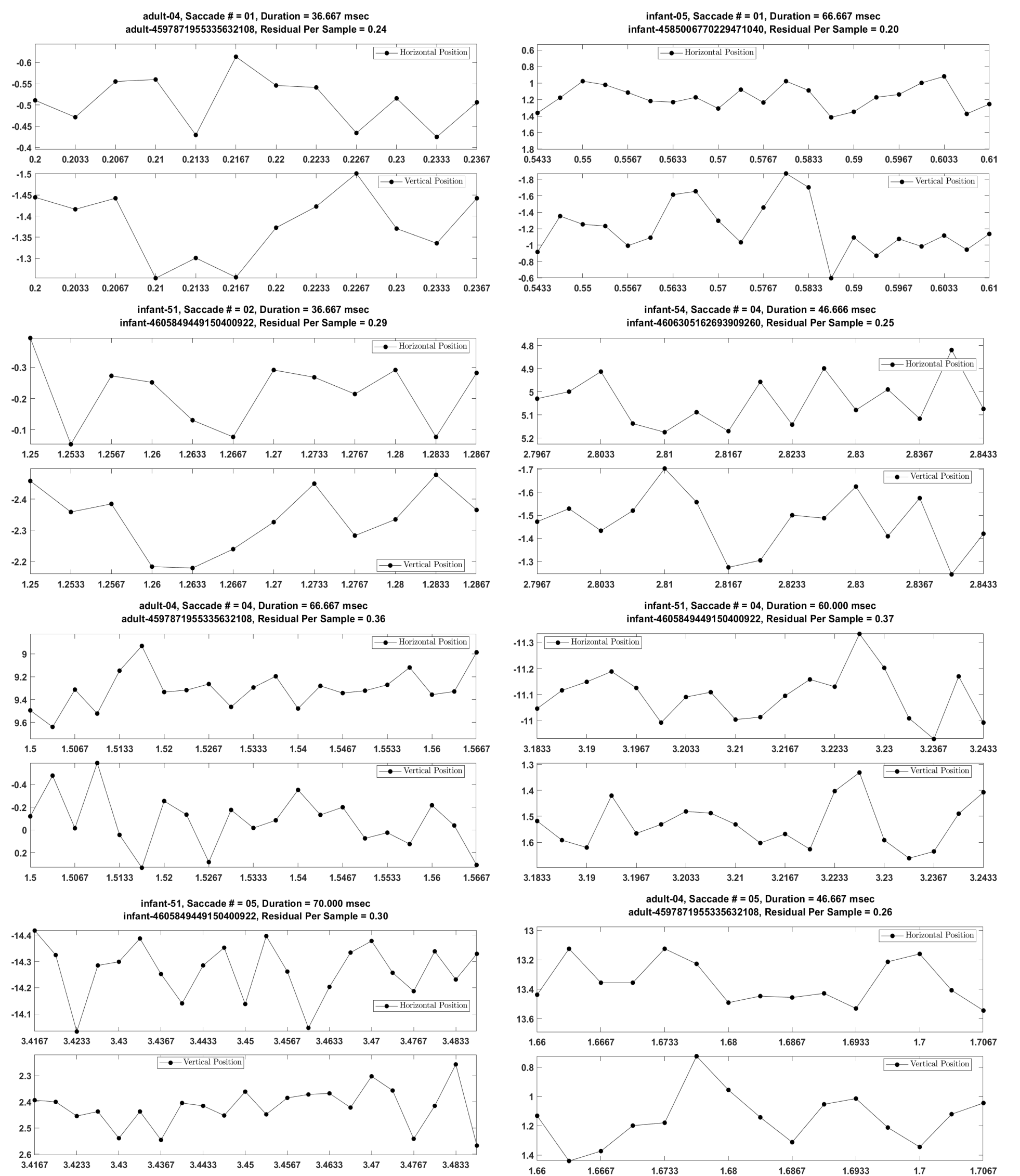}}
\begin{picture}(400,20)(0,0) \put(-4,5){{\parbox{500\unitlength}{
\textbf{Fig. 12: Eight Examples of ``bad'' saccades}}
}}
\end{picture}

\frame{\includegraphics[width=1.0\textwidth]{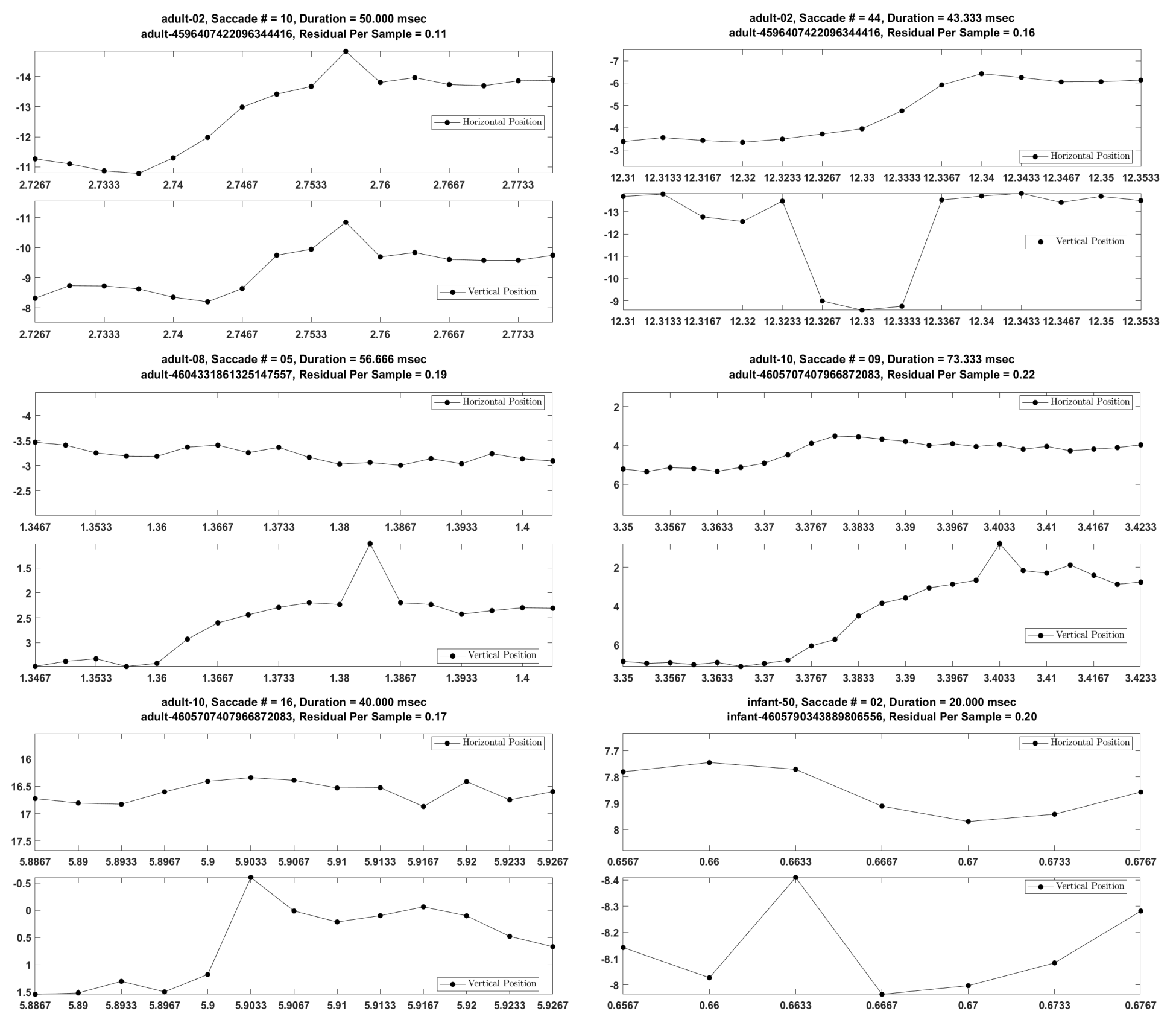}}
\begin{picture}(500,35)(0,0) \put(0,6){{\parbox{500\unitlength}{
\textbf{Fig.13: Six examples of saccades candidates. The amplitude of these saccades would be incorrect if determined by taking the maximum and minimum in the inter-fixation interval.}}
}}
\end{picture}

\subsection{Mismarks}
There are a number of ``fixations'', as indicated by coder MN (the only coder examined) that contain obvious saccades (Figures 14, 15, and 16).  No eye movement expert would ever intentionally include such saccades in fixations.  There must be some kind of mistake.  To find these I found the maximum change in the horizontal direction and the vertical direction during each fixation.  I then expressed this as a percentage of the range of each signal (Hor: 0-1920, Ver: 0-1080). Fixations with any missing data were not included in this analysis. For one study in particular (Figure 10), 21 of 37 fixations had percent changes greater than 5\% (23/37).  The majority of “fixations” marked in this dataset have obvious saccades in them (Figure 16).


\quad\quad\quad\quad\quad\quad\quad\quad\quad\quad\quad
\frame{\includegraphics[width=0.5\textwidth]{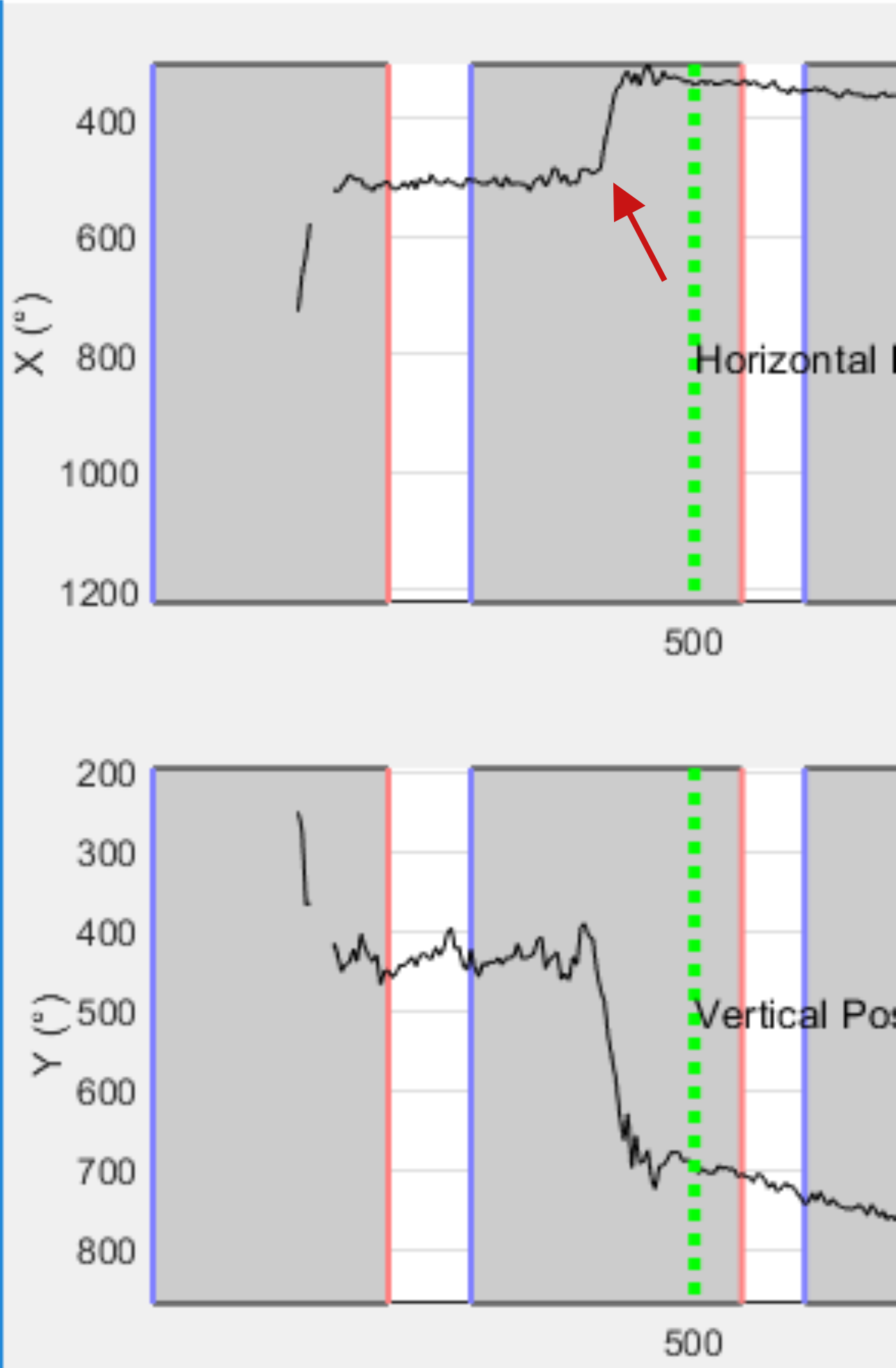}}
\begin{picture}(500,60)(0,0) \put(-,5){{\parbox{500\unitlength}{
\textbf{Figure 14. Eye movement signals for infant data set ``infant-4606305162693909260''.}  This figure is a cropping of the data figures as drawn by the code distributed with the data (see footnote 1).  Although the plots are labeled in units of degrees, they are, in fact, in units of pixels.  Fixation periods are in grey.  The top plot is the horizontal signal and the bottom plot is the vertical signal.  The dashed green vertical line marks the 500 msec time point (a feature I added to the plots).  The saccade amplitude for the saccade indicated by the arrow is 11.2\% of the horizontal range and 31\% of the vertical range. (Note that the entire potential signal range is not shown.). There are 3 other fixations in this recording that contain obvious saccades.}
}}
\end{picture}

\quad\quad\quad\quad\quad\quad\quad\quad\quad\quad
\center
\frame{\includegraphics[width=0.8\textwidth]{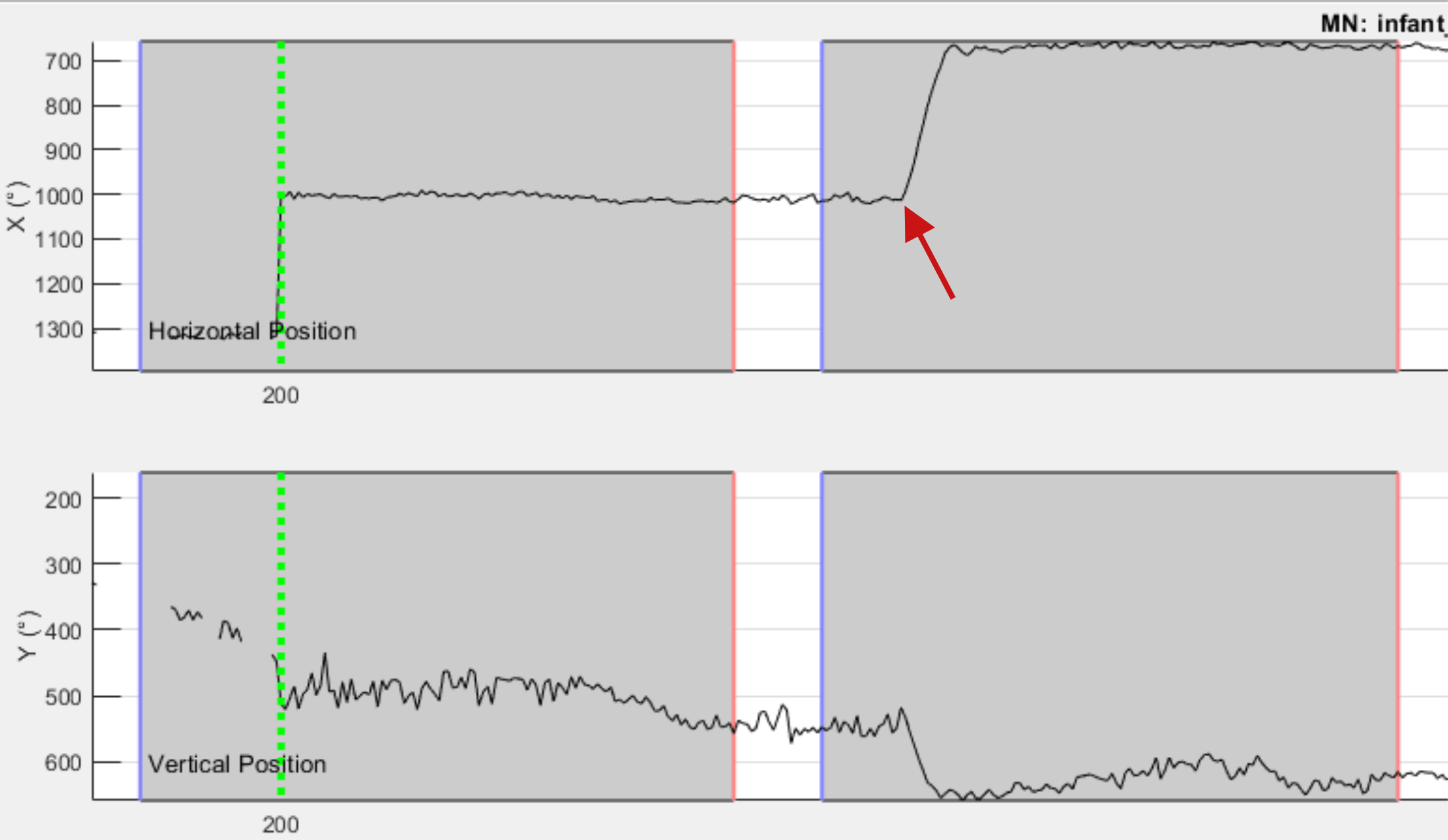}}
\begin{picture}(500,20)(0,0) \put(0,5){{\parbox{500\unitlength}{
\textbf{\\Figure 15: Infant data set ``infant-4585006770229471040''.} See caption for Figure 8. The saccade amplitude for the saccade indicated by the arrow is 18.9\% of the horizontal range and 13.1\% of the vertical range.}
}}
\end{picture}

\vspace{5mm}
\quad\quad\quad\quad
\centering
\frame{\includegraphics[width=0.8\textwidth]{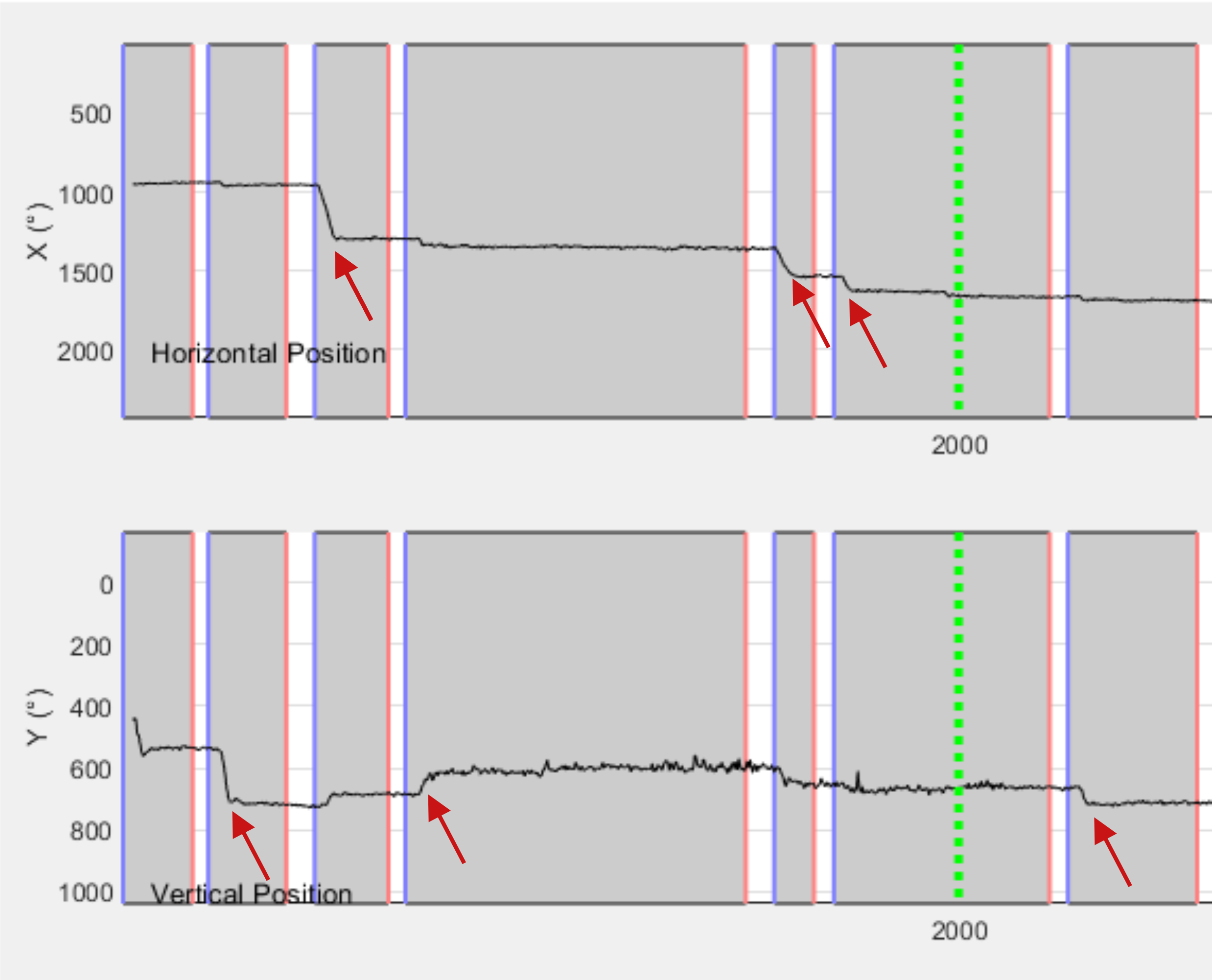}}
\begin{picture}(500,25)(0,0) \put(0,5){{\parbox{500\unitlength}{
\textbf{Figure 16: Signals from adult data set `` adult-4597871955335632108''.}  The first 7 ``fixations'' are shown. The 6 arrows indicate obvious saccades.}
}}
\end{picture}

\justifying
\section{Discussion}
Note that if the Hooge et al. (2018) \cite{HoogeF1} paper had been titled: ``Is human classification by experienced untrained observers a gold standard in fixation detection when data quality is very poor, data are error-filled, data presentation was not optimal, and the analysis was seriously flawed?'', I would have no case to make.  But it was not titled that way. It was titled without qualification.  In the absence of evidence to the contrary, it is reasonable to assume that some studies may provide evidence that human raters are a gold standard and other studies may not.  None of this conditionality appears to be recognized by Hooge et al. \cite{HoogeF1}, at least as far as their title is concerned.  The title appears as a blanket statement.

\subsection{Data of low quality}
Hooge et al. \cite{HoogeF1} based their study on recordings from a relatively low-quality recording device\footnote{For example, in a study of system precision \cite{Precision} the Tobii TX 300 ranked \nth{8} of 12 systems}. They included 60 infant recordings, many with low quality. For example, 11 of the 60 infant recordings had more than 50\% missing data, and one infant had 94\% missing data. Also, in their Table 2, they indicate that for every RMS noise estimate, RMS error was higher in infants than adults.   It should be noted that Hooge et al. \cite{HoogeF1} did not use any forehead or chinrest during data collection.  This is understandable in the case of infants. But we consider it axiomatic that, all other things being equal, eye-movement recordings with head restrained subjects will be higher quality than eye-movement recordings without head restraint/support.

\subsection{Erroneous data}
In examining the data made available by Hooge et al. \cite{HoogeF1} we found a number of fixations that were marked by experts that clearly had saccades in them.  No fully conscious and attentive eye movement expert would ever include such obvious saccades in fixation periods.  We also found cases where fixations were classified when there was no signal at all.  

\subsection{Coders not presented with all the data}
We provided evidence that perfectly informative eye movement signals were not presented to coders when making classification decisions.  Also, Hooge et al. present their data in pixel units rather than the more informative degrees of visual angle and velocity in units of deg/sec.  We believe that knowing the size of movements in degrees can assist human raters in making more accurate judgements, although we provide no proof of this statement at this time.  We see no advantage to using pixels over degrees. 

\subsection{Data analysis seriously flawed}
Long stretches of missing data were included as if properly labelled by human raters as either fixation or non-fixation.  In some cases, an event was classified as either fixation or non-fixation when the majority of the event or the entire event consisted of missing data.  We take it as axiomatic that human raters cannot classify datastreams into fixation and non-fixation when there is no signal. 

Hooge et al. \cite{HoogeF1} count as saccades many inter-fixation intervals that simply contain no saccade.  They also they measure saccade amplitude from these same, "no-saccade" sections of data.  In additon, they measure the amplitude of saccades in sections of data where these estimates are likely to be incorrect.  

\subsection{Conclusion}
If one's goal is to make a general statement that, in all cases, human classification of fixation is not a gold standard, human raters must be given the very best chance to be gold standards.  The recording quality must be very high, the interface must be optimized and the analysis scrupulously performed.  Otherwise, researchers can always say that humans might have performed better under better circumstances.  We take it as axiomatic that the better the quality of the data, the better human classifiers will perform.  A high quality study would minimally require that head restraint of some sort be employed, that signals be calibrated into degrees of visual angle and that only normal adults are recorded.  Unusual populations such as infants should be avoided.  Such a study would transparently display examples of recordings so that readers can readily determine signal quality. A reasonable upper limit on the percentage of each recording that consists of missing data or artifact should be employed.  Furthermore, the interface presented to the coders must be optimized for them. This will require a cycle of trial and error with attempts to classify recordings and consensus discussions comparing various interface designs.  Certainly it would be important to present to the coders all the relevant signals.  The analysis of agreement must also be optimal.  If one is going to make a statement about the agreement between raters on events like saccades, it is best to have the coders actually code the start and end of the saccades rather than to define saccades as unexamined inter-fixation intervals.  It does not make sense to include events, the majority of which, or the entirety of which consists of missing data.  For event-level matching, maximum event overlap or intersection-over-union metrics seem to be a better choices than the unjustified matching of the comparison event occurring earliest in time.

\section{Acknowledgements}
I would like to express my gratitude to Henry K. Griffith, Ph.D. for his incredibly careful proofreading of an earlier version of this manuscript.

\section*{Conflict of interest}
I have no conflict of interest.

\end{document}